\documentclass[a4paper,11pt]{article}
\usepackage{pos}

\usepackage{amssymb,amsmath,amsfonts}
\usepackage{mathtools}
\usepackage{mathrsfs}
\usepackage{bbm}
\usepackage{slashed}
\usepackage{nicefrac}

\usepackage{graphicx}

\usepackage{simplewick}

\usepackage{hyperref}
\usepackage{xparse}
\usepackage{xspace}

\usepackage{tikz}
\usetikzlibrary{decorations.pathmorphing}
\usetikzlibrary{automata,positioning}

\usepackage{cancel}
\usepackage[normalem]{ulem}

\usepackage{xifthen}
\usepackage{dsfont}
\usepackage[titletoc]{appendix}
\usepackage{booktabs}
\usepackage{units}

\newcommand{\gettitle}{}
\hypersetup{linkcolor=black
	colorlinks,
	linkcolor={red!75!black},
	citecolor={blue!75!black},
	urlcolor={blue!75!black},
	pdftitle={\gettitle},
	pdfauthor={Boussarie, Mehtar-Tani},
	pdfkeywords={Perturbative QCD} {Small-x}{TMD},
	bookmarksopen=true,
	bookmarksopenlevel=2,
	bookmarksnumbered=true
}
\makeatletter
\newcommand\makebig[2]{%
  \@xp\newcommand\@xp*\csname#1\endcsname{\bBigg@{#2}}%
  \@xp\newcommand\@xp*\csname#1l\endcsname{\@xp\mathopen\csname#1\endcsname}%
  \@xp\newcommand\@xp*\csname#1r\endcsname{\@xp\mathclose\csname#1\endcsname}%
}
\makeatother
\makebig{biggg} {1.3}

\def\del{\partial}

\newcommand{\eqn}[1]{Eq.~\eqref{#1}}

\long\def\comment#1{ }

\def\be{\begin{eqnarray*}}
\def\ee{\end{eqnarray*}}
\def\beq{\begin{eqnarray}}
\def\eeq{\end{eqnarray}}
\newcommand{\bea}{\beq \begin{aligned}}
\newcommand{\eea}{\end{aligned}\eeq}

\def\A{{\boldsymbol A}}
\def\x{{\boldsymbol x}}

\def\z{{\boldsymbol z}}

\def\0{{\boldsymbol 0}}
\def\l{{\boldsymbol l}}

\def\x{{\boldsymbol x}}

\def\z{{\boldsymbol z}}

\def\A{{\boldsymbol A}}

\def\A{{\boldsymbol A}}

\def\rmd{{\rm d}}

\def\and{ \quad\text{and}\quad}

\def\cP{{\cal P}}

\usepackage{color}
\definecolor{rbcolor}{rgb}{0.7,0.1,0}

\newcommand\rbout{\marginpar{\color{rbcolor}$\clubsuit$}\bgroup\markoverwith{\color{rbcolor}{\rule[0.4ex]{2pt}{0.8pt}}}\ULon}

\definecolor{ymtcolor}{rgb}{0.1,0,0.7}

\newcommand\ymtout{\marginpar{\color{ymtcolor}$\clubsuit$}\bgroup\markoverwith{\color{ymtcolor}{\rule[0.4ex]{2pt}{0.8pt}}}\ULon}

\title{On gauge invariance of transverse momentum dependent distributions at small-x}

\author[a,b]{Renaud Boussarie}
\author*[a,c]{Yacine Mehtar-Tani}

\affiliation[a]{Physics Department, Brookhaven National Laboratory, Upton, NY 11973, USA}
\affiliation[b]{Los Alamos National Laboratory, Mail Stop B283, Los Alamos, NM 87545, USA}
\affiliation[c]{RIKEN Research Center, Brookhaven National Laboratory, Upton, NY 11973, USA}

\emailAdd{renaud@lanl.gov}
\emailAdd{mehtartani@bnl.gov}

\abstract{Transverse momentum dependent (TMD) distributions at small x exhibit a rich infinite twist structure that encompasses the leading twist (partonic) distributions as well as the physics of gluon saturation. Progress to further the connection between the standard TMD framework at moderate x and small x has been recently made.  In this context, we show that light cone Wilson line operators at small-x can be formulated in terms of transverse gauge links. This new formulation of small x operators allows a direct matching with the standard leading twist gluon TMD distributions and provides an efficient and general prescription for computing TMD distributions at small x beyond leading twist. 

}

\FullConference{%
  HardProbes2020\\
  1-6 June 2020\\
  Austin, Texas}

\begin{document}
\maketitle

\section{Introduction\label{sec:intro}}
The partonic picture of the proton arises naturally in the Bjorken limit, where the hard scale $Q$ and center-of-mass energy $s$ tend to infinity, while $x \sim Q^2/s\sim 1$. Long distance physics where confining forces take place is then encoded into parton distribution functions that factorize from the hard matrix elements that only involves weekly interacting partons. In addition to the Bjorken limit, there exists another regime of QCD, the so-called Regge limit that is expected to be reached at high enough energy and small $x\sim Q^2/s \to 0$. In this regime, where the gluon number increases rapidly, non-linear recombination effects become important causing the gluon number density to saturate in the proton. In this high density environment, partons are no longer the pertinent degrees of freedom and the physics is best described by strong classical fields $A\sim 1/g$ which are resummed into light cone Wilson lines acting on hadronic states ~\cite{McLerran:1993ni,Balitsky:1995ub,Balitsky:1998kc,JalilianMarian:1997gr,Iancu:2000hn}.  

In the past few years, a lot of progress has been made in relating small-x physics to transverse momentum dependent (TMD) distributions that have been thoroughly studied in the Bjorken limit.  In~\citep{Dominguez:2011wm}, the authors showed how to extract the small $x$ limit of a Transverse Momentum Dependent (TMD) distribution by performing a transverse gradient expansion of Wilson line operators in the context of dijet production in the forward direction. This twist expansion yields two field strength tensors $F^{-i}$ at leading twist describing the exchange of two physically polarized gluons albeit multiplied by gauge links to ensure gauge invariance. Although this small $x$ calculation was performed in the ``wrong'' light cone gauge, the final result is manifestly gauge invariant. This suggests that the reformulation of small x observables in an explicitly gauge invariant way from the get-go would permit a more transparent connection with the standard TMD framework. 

There have been few attempts to unify small and moderate $x$ evolution equations for leading twist TMD's recently~\citep{Balitsky:2015qba, Balitsky:2019ayf}. However, no equivalence was formulated beyond the leading power in $k_\perp/Q$ until very recently. In~\citep{Altinoluk:2019fui,Altinoluk:2019wyu}, it was shown that a class of observables at small $x$ could be rewritten as the eikonal limit of
an infinite twist TMD framework. This new formulation of small-$x$ physics was based on a reorganization of the twist expansion. 
Although the final expressions were fairly simple it would be cumbersome to generalize to other classes of observables.

In this work \cite{Boussarie:2020vzf}, we revisit the aforementioned equivalence, and uncover the underlying geometric structure which preserves the explicit gauge invariance of the operators. For this purpose,
we demonstrate that pairs of Wilson line operators have a transparent and powerful 
formulation in terms of transverse gauge links. In this new approach for TMD's at small-$x$ in terms of transverse gauge link operators it will be straightforward  to generalize to other observables.

\section{Transverse parallel transport and the dipole operator in DIS dijet production}
Before discussing sub-amplitudes involved in computing small x observables, let us first show how transverse gauge links can naturally arise from light cone Wilson lines. Transverse gauge links constitute the corner stone of our new formulation of small x  in terms of TMD distributions. 

In the semi-classical formulations of small $x$ physics it is common to computes scattering amplitudes in the ``wrong'' light cone gauge, i.e., $A^+=0$ for a left moving target, by fixing the residual gauge freedom such that the classical transverse field vanishes and $x^+$ infinities. Therefore, only the $-$ component of the background field is taken into account through light-like Wilson lines along the $x^+$ direction. On the other hand, in the light-cone gauge $A^-=0$, the target gauge field is a transverse pure gauge $
A^i =\frac{i}{g}\Omega \del^i \Omega^{-1}\, $, where $\Omega$ is an element of SU(3). 

The connection to TMD physics requires the formulation of the problem in terms of field strength tensors, typically introduced by performing a gradient expansion in transverse position space that yields $\partial^i A^- \sim F^{i-}$. In effect, this corresponds to the parallel transport of Wilson line operators on the transverse plane as will be shown shortly. For any pair of transverse positions (\textbf{$\boldsymbol{x}_{1},\boldsymbol{x}_{2}$}) and
defining $\boldsymbol{r}=\boldsymbol{x}_{1}-\boldsymbol{x}_{2}$, we can readily write 
$\Omega_{\boldsymbol{x}_{1}}(x^+)
=\Omega_{\boldsymbol{x}_{2}}(x^+)-r^{i}\int_{0}^{1}\rmd s \, (\partial^{i}\Omega)_{\boldsymbol{x}_{2}+s\boldsymbol{r}}(x^+)$. 
Now, note that $\partial^{i}\Omega_{\boldsymbol{x}}(x^+)=ig {A}^{i}(x^+,\boldsymbol{x})\Omega_{\boldsymbol{x}}(x^+)\label{eq:igA}$
where ${A}^{i}$ (with $i=1,2$)  is the target pure gauge field. Combining
the two remarks above then multiplying by $\Omega_{\boldsymbol{x}_{2}}^{-1}(x^+)$
on the right yields $\Omega_{\boldsymbol{x}_{1}}(x^+)\Omega_{\boldsymbol{x}_{2}}^{-1}(x^+)=1-ig \, r^{i}\int_{0}^{1} \rmd s\, A^i(x^+,\boldsymbol{x}_{2}+s\boldsymbol{r})\, \Omega_{\boldsymbol{x}_{2}+s\boldsymbol{r}}(x^+)\Omega_{\boldsymbol{x}_{2}}^{-1}(x^+)$. Notice that  the latter equation defines a Wilson line along the straight line trajectory parametrized by the real number $s$ with values between 0 and 1 and such that $\boldsymbol{z}(0)=\boldsymbol{x}_2$ and $\boldsymbol{z}(1)=\boldsymbol{x}_1$, that is $\boldsymbol{z}(s)=\boldsymbol{x}_{2}+s\boldsymbol{r}$. 
This Wilson line, that we denote as $\Omega_{\boldsymbol{x}_{1}}(x^+)\Omega_{\boldsymbol{x}_{2}}^{-1}(x^+)\equiv [\boldsymbol{x}_1,\boldsymbol{x}_2]_{x^+}$, solves the following equation:
\begin{equation} 
[\boldsymbol{x}_{1},\boldsymbol{x}_{2}]_{x^+}=1-ig\!\int_{\boldsymbol{x}_{2}}^{\boldsymbol{x}_{1}}\!\!\rmd \z\cdot\A(x^+,\boldsymbol{z})[\boldsymbol{z},\boldsymbol{x}_{2}]_{x^+}\,.\label{eq:linkdefL}
\end{equation}
Although this transverse gauge link was constructed for a straight line trajectory it can be easily shown that it is independent of the trajectory connecting the endpoints so long as the transverse field is a pure gauge. \\

\begin{figure}[]
\begin{center}
\includegraphics[width=0.25\linewidth]{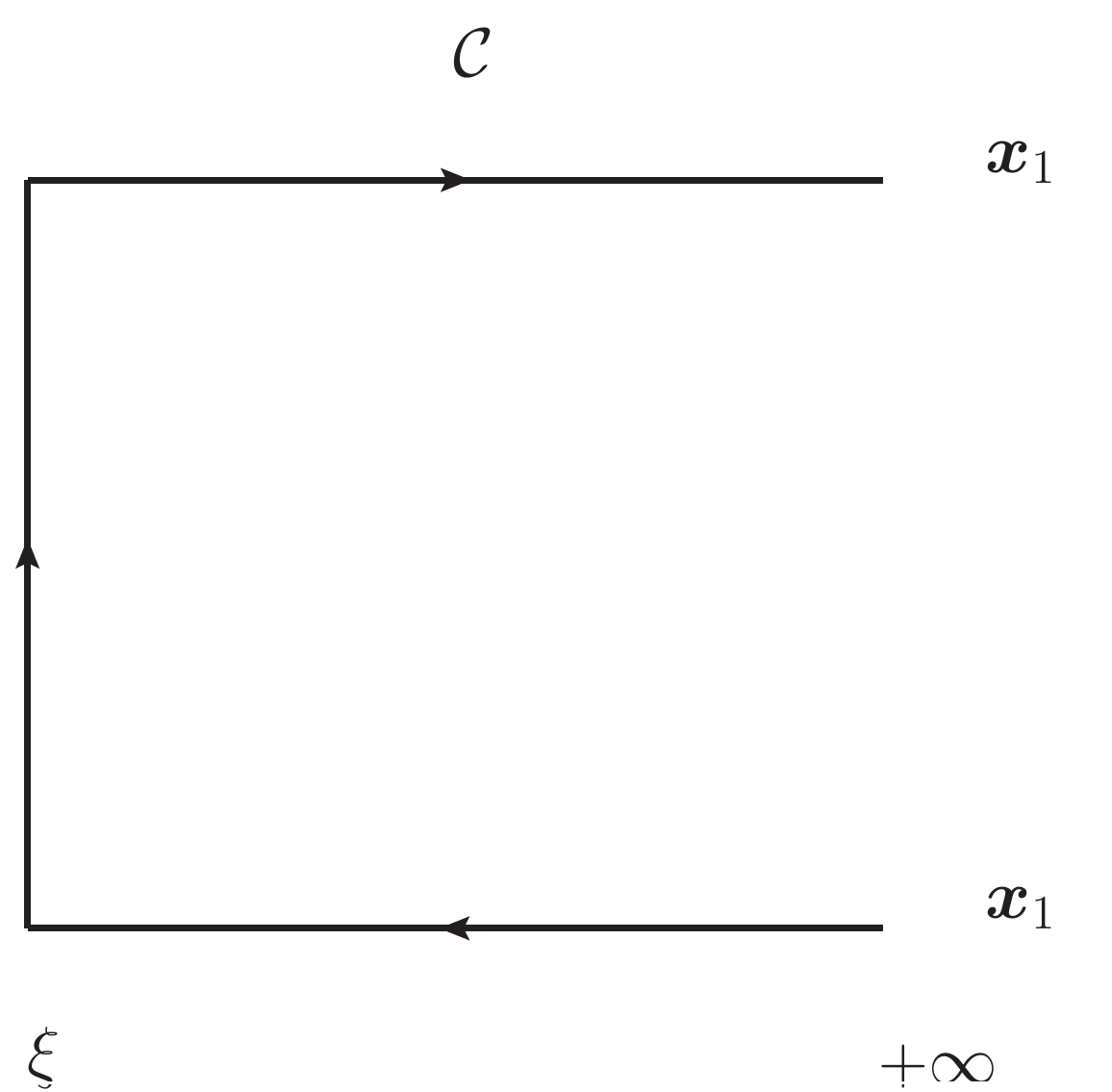} \qquad\qquad\includegraphics[width=0.4\linewidth]{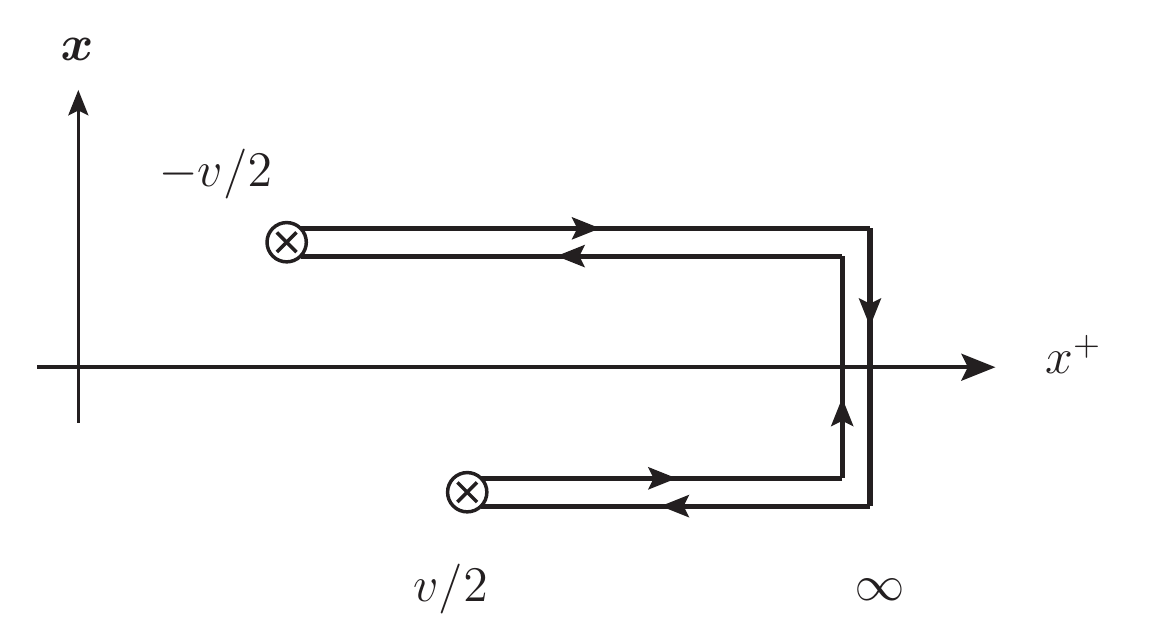}
\caption{Diagrammatic representation of the dipole operator \eqn{eq:dipoleinv} (left). Gauge link structure of the WW distribution (right).  \label{fig:WWlinks}}
\end{center}
\end{figure}

Now let us consider the (non-singlet) dipole operator that one encounters in dijet production in eA collisions where the target is  most commonly described by the $A^-$ gauge field,
\begin{equation}
{\cal M}_{ij}\sim \left(U_{\boldsymbol{x}_{1}}(\infty,\xi)U_{\boldsymbol{x}_{2}}^{\dagger}(\xi,\infty)\right)_{ij}\,.\label{eq:dipole}
\end{equation}
where $  U_{\boldsymbol{x}}(\infty,\xi) =\cP \exp\left[ig \int_{\xi }^\infty \rmd x^+\, A^-(x^+, \x) \right]$, 
is a path ordered Wilson line along the $x^+$ direction. Obviously, \eqn{eq:dipole} is not manifestly gauge covariant. To remedy this let us perform a gauge rotation of $A^-$ field $
A^{-}(x^{+},\boldsymbol{x})  \rightarrow\Omega_{\boldsymbol{x}}(x^{+})A^{-}(x^{+},\boldsymbol{x})\Omega_{\boldsymbol{x}}^{-1}(x^{+})-\frac{1}{ig}\Omega_{\boldsymbol{x}}(x^{+})\partial^{-}\Omega_{\boldsymbol{x}}^{-1}(x^{+})$.
Under such transformation the Wilson line yields $ U_{\boldsymbol{x}}(\infty,\xi) \to \Omega_{\boldsymbol{x}}(\infty) U_{\boldsymbol{x}}(\infty,\xi) \Omega_{\boldsymbol{x}}^{-1}(\xi)$.  Assuming that the gauge field vanishes at light-cone infinity for simplicity, one is left with the following structure (see Fig.~\ref{fig:WWlinks} (left))
\beq{\cal M}\to  U_{\boldsymbol{x}_{1}}(\infty,\xi)[\boldsymbol{x}_{1},\boldsymbol{x}_{2}]_\xi U_{\boldsymbol{x}_{2}}^{\dagger}(\xi,\infty)\,, \label{eq:dipoleinv}
\eeq
where we have used the transverse gauge link derived previously which connects the quark and the antiquark at $\xi$.  
A diagrammatic depiction of $\mathcal{O}_{\xi}(\boldsymbol{x}_{1},\boldsymbol{x}_{2})$  is given in Fig.~\ref{fig:WWlinks}, left panel. 
In Eq.~(\ref{eq:dipoleinv}), it is possible to absorb the longitudinal Wilson lines into the transverse link with the
following change of variables
${A}^{i}(\xi,\boldsymbol{z})\rightarrow\hat{A}^{i}(\xi,\boldsymbol{z})\equiv U_{\boldsymbol{z}}(\xi,\infty){A}^{i}(\xi,\boldsymbol{z})U_{\boldsymbol{z}}^{\dagger}(\xi,\infty)+\frac{1}{ig}(\partial^{i}U_{\boldsymbol{z}})(\infty,\xi)U^\dag_{\boldsymbol{z}}(\xi,\infty)$. We have now established that $
{ \cal M} \sim  [\hat{\boldsymbol{x}}_{1},\hat{\boldsymbol{x}}_{2}]_{\xi}$ 
which allows to understand the dipole operator as a transverse string
built from the rotated (hatted) transverse gluon fields.  \\

As a simple application (for more details see \cite{Boussarie:2020vzf})  let us now briefly show that given the above reformulation of the dipole operator it is straightforward to recover the  Weizs\"{a}cker-Williams (WW)  TMD.
By iterating  (\ref{eq:linkdefL}) once, one finds
\begin{align}
&{ \cal M}\sim 1+ig\!\int_{\boldsymbol{x}_{2}}^{\boldsymbol{x}_{1}}\!\!\rmd\z^{i}\hat{\boldsymbol{A}}^{i}(\xi,\boldsymbol{z})+(ig)^{2}\!\int_{\boldsymbol{x}_{2}}^{\boldsymbol{x}_{1}}\!\!\rmd \z^{i}\!\int_{\boldsymbol{x}_{2}}^{\boldsymbol{z}}\!\!\rmd \z^{\prime j}\hat{\boldsymbol{A}}^{i}(\xi,\boldsymbol{z})[\hat{\boldsymbol{z}},\hat{\boldsymbol{z}}^{\prime}]_{\xi}\hat{\boldsymbol{A}}^{j}(\xi,\boldsymbol{z}^{\prime})\,.
\end{align}
The second and third terms correspond to the one-body and two-body amplitudes, respectively,  derived in~\citep{Altinoluk:2019wyu} and we have shown that they emerge from a transverse link where the leading genuine twist $\propto A^i$, corresponding to the exchange of a single (transversely polarized) gluon,  is separated form higher twist. Focusing on the former contribution, we readily obtain, upon multiplication by the complexe conjugate amplitude, 
\beq
|{ \cal M}|^2 \sim \langle P| F^{i-}(\frac{v}{2}) \,\mathcal{U}_{\frac{v}{2},-\frac{v}{2}}^{\left[+\right]}\,F^{j-}(-\frac{v}{2})\,\mathcal{U}_{-\frac{v}{2},\frac{v}{2}}^{\left[+\right]},|P  \rangle \,.
\eeq
where we have used $\hat A^{i}(\xi,\boldsymbol{z})=\int_{\xi}^{\infty}\!\!\rmd z^+\,U_{\boldsymbol{z}}(\infty,z^+)F^{i-}(z^+,\boldsymbol{z})U_{\boldsymbol{z}}^{\dagger}(z^+,\infty)$ and $\mathcal{U}^{\left[+\right]}$'s are "staple" shaped gauge link structures involving future pointing light cone Wilson lines and we have set $ z=v/2$ and $-v/2$  in amplitude and its complex conjugate respectevely.  Here we recognize the field strength and gauge link structure of the WW gluon distribution, depicted in Figure~\ref{fig:WWlinks}, that appears in the improved-TMD framework \cite{vanHameren:2016ftb} and in the all twist formulation derived in Ref.~\cite{Altinoluk:2019wyu}

\section{Extension to generic color structures and multiple-line operators}\label{sec:gen}
A general prescription for arbitrary color structure and multiple line operators can easily be inferred from the above dipole operator, based on re-expressing the light cone Wilson lines in terms of transverse gauge links. More details can be found in the original article \cite{Boussarie:2020vzf} which these proceedings are based upon. Besides the technical benefits of this approach that provides a powerful tool to extend the TMD/small-x equivalence \cite{Altinoluk:2019fui,Altinoluk:2019wyu}, it also allows for a new geometric interpretation of hadronic operators at small x. Finally, it will be  interesting to investigate implications beyond tree level and possible connections to the Bjorken limit.

\section*{Acknowledgements} 
This work was supported by the U.S. Department of Energy, Office of Science, Office of Nuclear Physics, under contract No. DE- SC0012704,
and in part by Laboratory Directed Research and Development (LDRD) funds from Brookhaven Science Associates.
Y. M.-T. acknowledges support from the RHIC Physics Fellow Program of the RIKEN BNLResearch Center. R. B. is supported by the LANL  Laboratory Directed Research and Development program.

\bibliographystyle{elsarticle-num}
\bibliography{MasterBibtex}

\end{document}